\documentclass[letterpaper]{article} 
\usepackage{aaai2026}  
\usepackage{times}  
\usepackage{helvet}  
\usepackage{courier}  
\usepackage[hyphens]{url}  
\usepackage{graphicx} 
\def\UrlFont{\rm}  
\usepackage{natbib}  
\usepackage{caption} 
\usepackage{amsfonts} 
\usepackage{amsmath}
\usepackage{amssymb}
\usepackage{booktabs} 
\usepackage{graphicx} 
\usepackage{algorithm}
\usepackage{algorithmic}

\usepackage{newfloat}
\usepackage{listings}
\usepackage{booktabs}    
\usepackage{array}        
\usepackage{graphicx}     
\usepackage{siunitx}     
\usepackage{caption}      

\newcolumntype{R}{r}  
\DeclareCaptionStyle{ruled}{labelfont=normalfont,labelsep=colon,strut=off} 
\floatstyle{ruled}
\newfloat{listing}{tb}{lst}{}
\floatname{listing}{Listing}
\title{Causal Reasoning Elicits Controllable 3D Scene Generation}

\author {
    Shen Chen\textsuperscript{\rm 1}, 
    Ruiyu Zhao\textsuperscript{\rm 2}, 
    Jiale Zhou\textsuperscript{\rm 2},  
    Zongkai Wu\textsuperscript{\rm 3}, 
    Jenq-Neng Hwang\textsuperscript{\rm 4}, 
    Lei Li\textsuperscript{\rm 4,5}\thanks{Corresponding author(lenny.lilei.cs@gmail.com)}
}
\affiliations {
    \textsuperscript{\rm 1}Zhejiang University \quad
    \textsuperscript{\rm 2}East China University of Science and Technology\\
    \textsuperscript{\rm 3}Skai Intelligence \quad
    \textsuperscript{\rm 4}University of Washington \quad
    \textsuperscript{\rm 5}VitaSight\\
}

\usepackage{bibentry}

\begin{document}

\maketitle

\begin{abstract}
Existing 3D scene generation methods often struggle to model the complex logical dependencies and physical constraints between objects, limiting their ability to adapt to dynamic and realistic environments. We propose CausalStruct, a novel framework that embeds causal reasoning into 3D scene generation. Utilizing large language models (LLMs), We construct causal graphs where nodes represent objects and attributes, while edges encode causal dependencies and physical constraints. 
CausalStruct iteratively refines the scene layout by enforcing causal order to determine the placement order of objects and applies causal intervention to adjust the spatial configuration according to physics-driven constraints, ensuring consistency with textual descriptions and real-world dynamics.
The refined scene causal graph informs subsequent optimization steps, employing a Proportional-Integral-Derivative(PID) controller to iteratively tune object scales and positions.
Our method uses text or images to guide object placement and layout in 3D scenes, with 3D Gaussian Splatting and Score Distillation Sampling improving shape accuracy and rendering stability.
Extensive experiments show that CausalStruct generates 3D scenes with enhanced logical coherence, realistic spatial interactions, and robust adaptability.
\end{abstract}

\begin{links}
    \link{Code}{https://causalstruct.github.io/}
\end{links}

\section{Introduction}
\label{sec:intro}

In recent years, 3D scene generation has advanced significantly in computer vision, graphics, and content creation. However, traditional methods still rely heavily on manual modeling and expert knowledge, making multi-object scene construction time-consuming and costly. Existing text-to-3D approaches have attempted to address this by using 2D diffusion models for optimizing 3D representations \cite{dreamgs,gsdreamer,scenescape} or employing 3D diffusion models for direct asset generation \cite{hong20243dtopia,zhou2025diffgs}. While these approaches have demonstrated success in synthesizing individual objects, they struggle with multi-object scene composition, often resulting in geometric distortion, spatial inconsistencies, and object drift. 

A well-structured spatial layout is essential for generating coherent 3D scenes, as it dictates object placement. Previous layout-generation approaches \cite{zhou2024layout,procedural-3D-LLM,layoutgpt,zheng2024editroom,yang2021text,li3d} rely on data-driven heuristics or LLM-based inference to ensure semantic and structural consistency. However, these methods primarily focus on static spatial constraint placement and overlook the interactions between objects on the overall scene. 

\begin{figure}[t]
    \centering
    \includegraphics[width=0.95\linewidth]{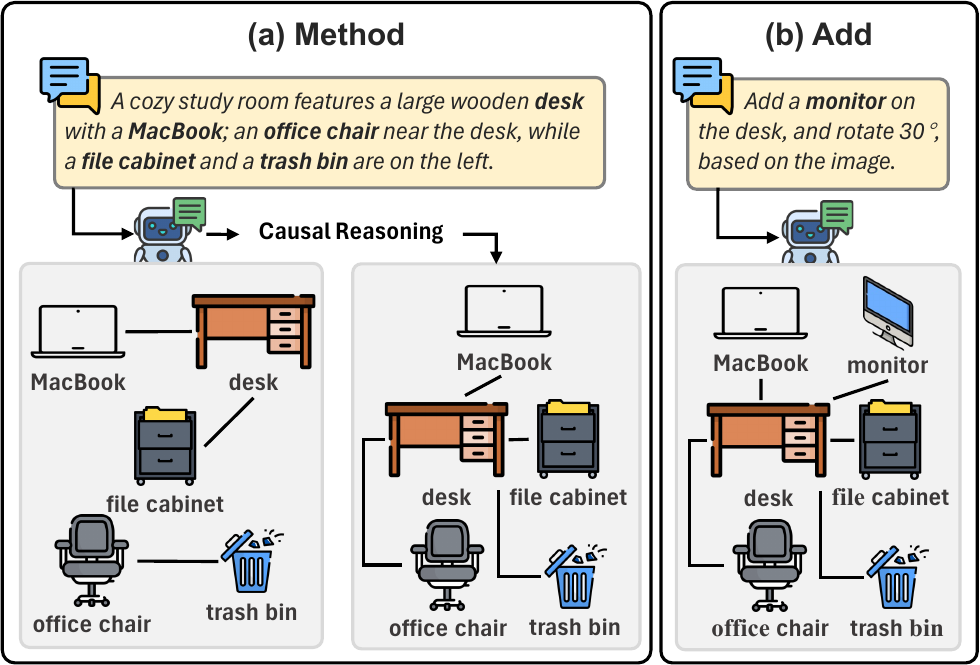}
    \caption{
        CausalStruct optimizes and controls 3D scene generation using either pure text 
        or a combination of text and image inputs, ensuring spatial coherence and 
        realistic object interactions through causal reasoning.
    }
    \label{fig:causalstruct}
\end{figure}

Causal reasoning in 3D scene generation establishes a hierarchy of directed dependencies among objects, ensuring their spatial and functional interactions adhere to real-world physical principles. Rooted in structural causal models (SCMs) \cite{causality}, it transcends statistical correlations by explicitly defining how the presence or state of one object causally influences the placement, orientation, or existence of others. Without causal reasoning, layout methods are unable to dynamically model object interactions, resulting in scenes with misaligned spatial relationships, improper functional placements, or floating objects that violate real-world physics.

To overcome these challenge, we propose CausalStruct, a novel framework that integrates causal reasoning into scene graph optimization. Using LLMs \cite{GPT4o, cai2025role}, we construct a scene graph where nodes represent objects and attributes, and edges encode relationships and physical dependencies. However, the LLM alone cannot accurately construct scene graphs due to its neglect of node properties and edge interactions, resulting in unrealistic scenes that defy physical laws.
Inspired by causal reasoning in structure discovery and relationship modeling \cite{kiciman2023causal,vashishtha2023causal}, we introduce a causal order mechanism to enforce logical sequencing in object placement. By computing a causal precedence through pairwise LLM reasoning, we ensure that objects follow physically consistent dependencies. Furthermore, to address uncertain or inconsistent edges, we compute the confidence of each edge using a Bayesian estimation \cite{cai2025bayesian} and determine whether to apply a causal intervention based on this confidence, where interventions on object states validate relationships through their physical impact on other scene elements, guiding whether to modify or retain the edge.

To refine the attributes of nodes (objects) in the causal scene graph and adjust them to realistic spatial proportions, we optimize attributes using a Multimodal Large Language Model (MLLM) \cite{GPT4o, li2024cpseg} that assesses spatial relationships. Each edge in the causal graph is assigned an attribute correction score, quantifying discrepancies in size and position. To ensure physically plausible adjustments, we apply a Proportional-Integral-Derivative (PID) controller \cite{willis1999proportional,crowe2005pid}, iteratively refining object attributes while maintaining scene stability. For enhanced geometric consistency and rendering stability, we integrate 3D Gaussian Splatting (3DGS) \cite{3dgs, 10890282} with Score Distillation Sampling (SDS) \cite{DreamFusion} at both the object and scene levels. Additionally, our method supports text-only and text-image combinations, enabling improved physical simulations and generating coherent 3D scenes.

\begin{itemize}
    \item We propose CausalStruct, integrating Causal Order, and Causal Intervention into scene graph optimization to enforce object dependencies and refine uncertain edges.
    \item We optimize whole 3D scene layout by refining object attributes with an MLLM and a PID controller.
    \item We enhance 3D scene generation by leveraging 3DGS with SDS to refine the Causal Scene, improving geometric consistency and rendering stability.
     \item We support both text-only and text-image combinations, enhancing scene generation and physical simulation accuracy.
\end{itemize}

\begin{figure*}[ht]
   \centering
   \includegraphics[width=0.9\linewidth]{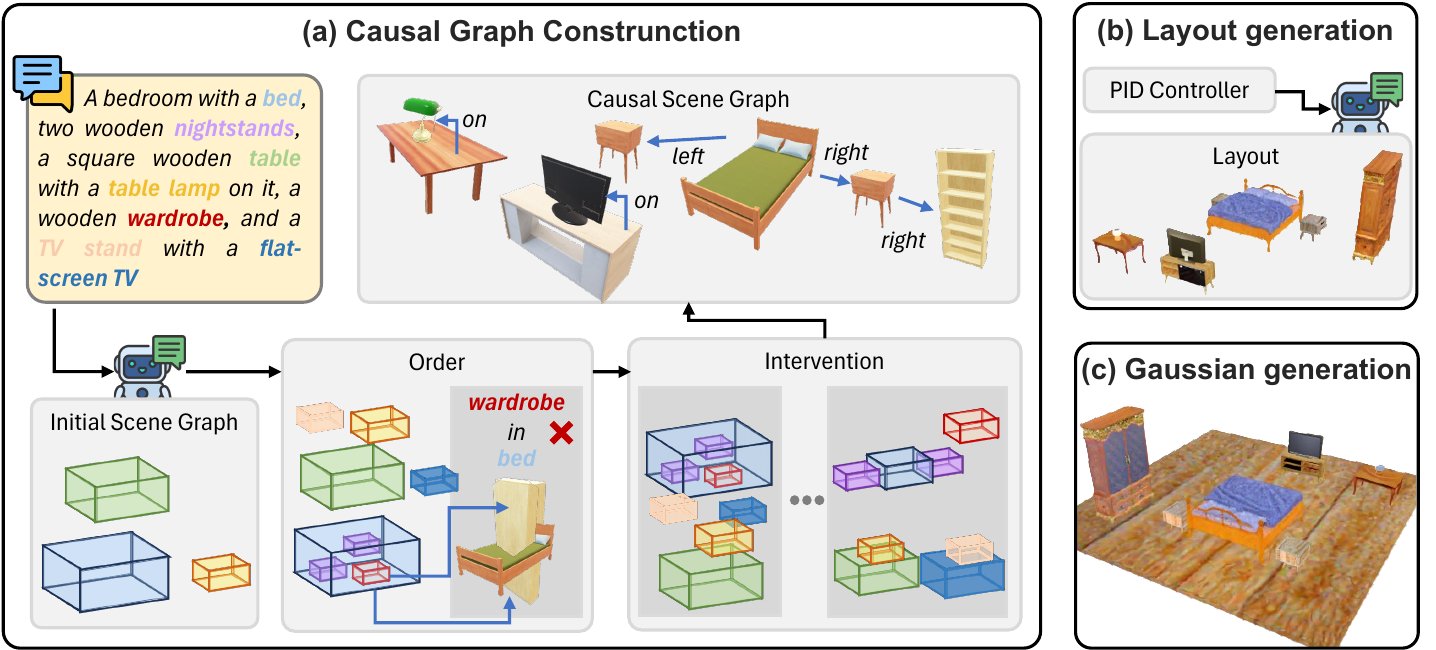}
   \caption{{Overview of our method.} Given a scene description, our method constructs a causal scene graph using LLMs and MLLMs with causal reasoning. A PID controller refines object scales and positions, ensuring spatial consistency. Additionally, objects and the scene are represented with 3D Gaussian Splatting  and optimized using Diffusion and SDS for high-fidelity rendering.}
   \label{fig:pipline}
 \end{figure*}

\section{Related Work}
\label{sec:related}


\paragraph{Text-Driven 3D Generation}\vspace{-0.5\baselineskip}
NeRF-based methods, such as DreamFusion \cite{DreamFusion} and Score Jacobian Chaining \cite{wang2023score}, leverage 2D diffusion models \cite{rombach2022high,saharia2022photorealistic} to synthesize single objects. Subsequent works, including Magic3D \cite{Magic3D}, Latent-NeRF \cite{metzer2023latent}, and 3DFuse \cite{seo2023let}, aim to enhance 3D generation quality under SDS constraints. ProlificDreamer \cite{ProlificDreamer} models 3D parameters as random variables, and introduces Variational Score Distillation (VSD) for improved optimization. While NeRF-based approaches effectively generate high-quality 3D objects, they suffer from inefficiency. To improve efficiency, 3DGS-based approaches \cite{dreamgs,gsdreamer} have been proposed for text-to-3D generation by integrating diffusion models with Gaussian splatting. Recent methods  \cite{zhou2025diffgs,GaussianCube,gsgen,jiang2024brightdreamer} utilize text-to-3DGS pipelines to facilitate object synthesis, achieving faster generation. While these methods enable diverse 3D generation from text prompts, they struggle to produce photorealistic multi-object scenes with complex geometry and high-fidelity textures due to their reliance on high-level semantic priors.

\paragraph{LLMs for Causal Discovery}\vspace{-0.5\baselineskip}
LLMs 
have significantly advanced causal inference by combining text-based dependency extraction with reasoning capabilities to uncover causal relationships.
LLMs enhance causal inference by performing pairwise causal reasoning to identify relationships between variables \cite{kiciman2023causal}. Beyond pairwise inference, LLMs contribute to causal graph construction, leveraging causal ordering to orient undirected edges \cite{vashishtha2023causal} and refining predictions through iterative feedback mechanisms \cite{ban2023query}. Additionally, LLMs enhance generalization by incorporating pretrained knowledge \cite{feng2024pre}, making them well-suited for capturing complex dependencies in structured reasoning. Since scene construction inherently involves causal relationships between objects, integrating LLMs with causal discovery into this process ensures coherent spatial layouts and semantically consistent object interactions.

\paragraph{Layout Generation}\vspace{-0.5\baselineskip}
Scene layout is fundamental to 3D scene generation, as it dictates the spatial arrangement, scale, and interactions of objects, directly influencing realism and coherence. 
Various methods have explored different strategies for scene composition and object placement. SceneSuggest \cite{scenesuggest} utilizes spatial constraints to infer supporting surfaces, while Physcene \cite{Physcene} and Text2nerf \cite{text2nerf} integrate diffusion models to enforce physically plausible layouts. Graph-based approaches have gained popularity in structuring object relationships and layout within scenes. PlanIT \cite{Planit} and SceneGraphNet \cite{scenegraphnet} encode spatial and functional dependencies, guiding object placement based on predefined constraints. GraphDreamer \cite{graphdreamer} and SceneWiz3D \cite{SceneWiz3D} incorporates LLMs with layout-based NeRF to further enhance scene composition, while GALA3D \cite{gala3d} and LayoutDreamer \cite{LAYOUTDREAMER} introduces layout-guided 3D Gaussian representation, leveraging adaptive constraints for geometry refinement and inter-object interactions. Despite their effectiveness in structuring layouts, these methods overlook causal dependencies in object placement and attributes, often leading to misalignment or floating objects. In this paper, we integrate causal reasoning to refine both object relationships and attributes, ensuring a more coherent and physically plausible scene.

\section{Methods}

As shown in Fig.~\ref{fig:pipline}, our method constructs causal-driven 3D Gaussian representations by integrating causal reasoning, PID-based optimization, and layout-guided representation. First, given a text description, we generate an initial scene graph using LLM and refine object relationships through causal reasoning. Second, to ensure spatial balance, PID Control optimizes object scale and position while preventing abrupt changes. Finally, Layout-Guided Representation builds the scene with 3DGS and optimizes it using Diffusion and SDS for spatial consistency and high-fidelity rendering.

\paragraph{Preliminaries}
3D Gaussian Splatting (3DGS) represents 3D scenes using anisotropic Gaussian primitives, denoted as $\{ \mathcal{G}_n \mid n = 1, \ldots, N \}$, with parameters including position $\mu_n \in \mathbb{R}^{3}$, covariance $\Sigma_n \in \mathbb{R}^{7}$, color $c_n \in \mathbb{R}^{3}$, and opacity $\alpha_n \in \mathbb{R}$. The Gaussian function is defined as:
\begin{equation}
     \mathcal{G}_n(p) = e^{-\frac{1}{2}(p-\mu_n)^T \Sigma^{-1}_n (p-\mu_n)},
\end{equation}
where $\Sigma_n$ is parameterized by a rotation matrix $R_n \in \mathbb{R}^4$ and a scaling matrix $S_n \in \mathbb{R}^3$.

For rendering, differential splatting projects the Gaussians onto camera planes, using a viewing transformation $W_n$ and the Jacobian matrix $J_n$ to obtain a transformed covariance. The color for a ray $r$ is computed as:
\begin{equation}
C_r(x) = \sum_{i \in M} c_i \sigma_i \prod_{j=1}^{i-1} (1 - \sigma_j), \quad \sigma_i = \alpha_i \mathcal{G}^{2D}_i(x).
\end{equation}
An adaptive density control mechanism dynamically adjusts the number of Gaussians to balance computational efficiency and scene detail.

\subsection{Causal Reasoning for Scene Graph}
\label{causal}

\paragraph{Causal Order for Scene Graph Optimization}
Ensuring a consistent and physically plausible scene layout, we define causal precedence among objects to establish a logical placement sequence.
Given a set of objects $ \mathcal{O} = \{o_1, o_2, ..., o_n\} $ and their spatial relations $ \mathcal{E} = \{e_{i,j} | o_i, o_j \in \mathcal{O}\} $, directly from a LLM, we define the causal order $ \prec $ as:
\begin{equation}
   o_i \prec o_j \iff C_{i,j} > C_{j,i},
\end{equation}
where $ o_i \to o_j $ represents a directed causal edge indicating that the placement of $ o_j $ depends on $ o_i $. $ C_{i,j} = \mathbb{P}(o_i \prec o_j \mid \text{LLM}(o_i, o_j)) $ represents the probability of $ o_i $ causally preceding $ o_j $. If $ C_{i,j} > C_{j,i} $, we adjust the scene such that $ o_i $ is placed before $ o_j $ in spatial reasoning, updating the edge set $ \mathcal{E} $ to the refined set $ \mathcal{E}_{\text{order}} $ by enforcing the inferred causal order.

\paragraph{Bayesian Edge Estimation}
The hallucinations of LLM lead to causal graphs containing false or physically unreasonable edges during linguistic reasoning.
We introduce Bayesian Edge Estimation to assess the confidence of each edge before Causal Intervention.

\textbf{Prior:} The prior probability $p(e_{i,j})$ of an edge $ e_{i,j} $ represents the inherent plausibility of the causal relationship between objects $ o_i $ and $ o_j $, before observing any additional evidence. Instead of assuming a uniform prior, we obtain the prior directly from LLM.

\textbf{Posterior:} For a candidate edge $ e_{i,j} $, its posterior probability given the Causal Order edges $ \mathcal{E}_{\text{order}} $ is computed using Bayes' rule:
\begin{equation}
    p(e_{i,j} \mid \mathcal{E}_{\text{order}}) = \frac{ p(\mathcal{E}_{\text{order}} \mid e_{i,j}) p(e_{i,j}) }{ p(\mathcal{E}_{\text{order}}) },
\end{equation}
where $ p(\mathcal{E}_{\text{order}} \mid e_{i,j}) $ is the likelihood that the LLM-generated edge set is observed given that $ e_{i,j} $ is correct, and $ p(\mathcal{E}_{\text{order}}) $ is the normalizing factor ensuring a valid probability distribution. We assume that the likelihood of $ \mathcal{E}_{\text{order}} $ given $ e_{i,j} $ is decomposable across all edges in $ \mathcal{E}_{\text{order}} $:
\begin{equation}
    p(\mathcal{E}_{\text{order}} \mid e_{i,j}) = \prod_{(o_i, o_j) \in \mathcal{E}_{\text{order}}} p(\text{LLM}(o_i, o_j) \mid e_{i,j}),
\end{equation}
where $p(\text{LLM}(o_i, o_j) \mid e_{i,j})$ represents the probability that the LLM correctly predicts the causal relationship between objects $ o_i $ and $ o_j $, obtained by querying the model multiple times and aggregating its responses.

\paragraph{Causal Intervention for Edge Refinement}

For edges with uncertain posterior probabilities, we conduct causal interventions using an MLLM to verify their correctness. Given two objects $ o_i $ and $ o_j $ connected by an edge $ e_{i,j} $, we evaluate the impact of modifying their spatial relationship on the entire scene. This intervention is performed by iterating over a set of candidate placements for the object and analyzing their effect using the rendered scene image. The MLLM determines whether each placement results in a physically and semantically plausible configuration.

To quantify the likelihood of each possible adjustment, we define the probability of $ D_{i,j} $ as:
\begin{equation}
    p(D_{i,j} \mid e_{i,j}) = 
    \sum_{r,s \in S} p(D_{i,j} = s \mid do(e_{i,j}=r)),
\end{equation}
where $ S $ is the set of all candidate placements for object $ o_j $, determined by a predefined position list. $do(e_{i,j}=r))$ means to interfere with edge $e_{i,j}$ to force its state to be set to $r$.
$ p(D_{i,j} = s \mid do(e_{i,j}=r)) $ represents the probability of state $ s $ being the modification for $ do(e_{i,j}=r)) $, which is estimated using an MLLM evaluation:
\begin{equation}
\begin{split}
p(D_{i,j} = s \mid do(e_{i,j}=r)) \\ = {}  \frac{1}{K}\sum_{k=1}^{K} 
& \mathbb{I}\Big(\text{MLLM}(o_i, o_j, I_{e_{i,j}}^{r}) \rightarrow s\Big),
\end{split}
\end{equation}
where $ \mathbb{I}(\cdot) $ is an indicator function that returns 1 if the MLLM selects state $ s $ for edge $ e_{i,j} $ in the $ k $-th trial, otherwise returns 0. $ I_{e_{i,j}}^{r} $ is the rendered scene image when object $ o_j $ is placed at candidate position $ r\in S $, and the MLLM evaluates whether this placement is reasonable. The intervention decision is determined by selecting the placement modification:
\begin{equation}
    s^* = \arg\max_{s \in S} p(D_{i,j} = s \mid do(e_{i,j}=r)) .
\end{equation}


\paragraph{Update Strategy}
Edges are classified into three categories based on their posterior probability and causal intervention results:
\begin{equation}
e_{i,j} =
\begin{cases} 
      e_{i,j} & p_{e_{i,j}} > \tau_1 \\
      s^* & p_{e_{i,j}} \leq \tau_1 \& p(\mathcal{E}_{\text{order}} \mid s^*) > \tau_2  \\
      \varnothing & p_{e_{i,j}} \leq \tau_1 \&  p(\mathcal{E}_{\text{order}} \mid s^*) \leq \tau_2
\end{cases},
\end{equation}
where $ \tau_1 $ and $ \tau_2 $ are confidence thresholds, $p_{e_{i,j}}$ represents $p(\mathcal{E}_{\text{order}} \mid e_{i,j})$, $p(\mathcal{E}_{\text{order}} \mid s^*)$ represents the posterior probability of $s^*$, and $\varnothing$ means remove the edge. This ensures that high-confidence edges remain, medium-confidence edges are validated through interventions, and low-confidence edges are discarded, leading to a reliable causal scene graph.

\subsection{PID Control Object Optimization}
\label{pid_controler}

Causal reasoning determines the initial placement of objects, while the PID controller fine-tunes their positions and scales to ensure physical plausibility and spatial accuracy. Each object pair (edge) in the scene graph is evaluated using an MLLM to accurately adjust edges, generating a scale correction score and position scores for precise spatial refinement.

We propose an optimization method based on a Proportional-Integral-Derivative (PID) controller to ensure proportional accuracy and visual balance in reconstructed scenes. 
The PID controller can effectively handle the non-independent relationships between edges by dynamically adjusting the proportional, integral, and derivative parameters to  to ensure that the objects in the edges do not overlap and ensure overall spatial coordination and accuracy.

The error signal, defined as the negation of the score, drives the PID controller to adjust the scale and position. The control signal $u$ is computed using the error $e$, accumulated error$\int e \, dt$, and change in error over time $\frac{de}{dt}$:
\begin{equation}
    u = K_p e + K_i \int e \, dt + K_d \frac{de}{dt},
\end{equation}
where $K_p$, $K_i$ and $K_d$ are the proportional, integral, and derivative coefficients. 

To implement PID control effectively, we introduce an actuator that converts the control signal $ u $ into practical adjustments of the scene's scale and position. 
The actuator ensures that adjustments remain within a predefined range, preventing instability or produce unrealistic results. 
The actuator output $ \Delta $ is formulated as a nonlinear function of $ u $, employing the hyperbolic tangent function to achieve smooth scaling:
\begin{equation}
    \Gamma = \Delta \cdot tanh(\frac{u}{\gamma}),
\end{equation}
where $ \Delta $ denotes the maximum permissible adjustment, and $ \gamma $ modulates the steepness of the response. This transformation constrains the output within the range $ -\Delta $ to $ \Delta $, ensuring smooth and controlled adjustments while preventing abrupt shifts in scale or position.

Once the nodes and edges are optimized, they are organized into a graph based on inter-object relationships, such as aligned spatial attributes and semantic associations. The LLM then directs the placement of these subgraphs by interpreting high-level prompts that specify the expected sizes, relationships, and spatial context of the objects. Additionally, based on the attribute values of each node, MLLM is utilized to finely adjust the scene graph, ensuring that the final structure accurately aligns with the textual description.

\begin{figure*}[ht]
   \centering
   \includegraphics[width=\linewidth]{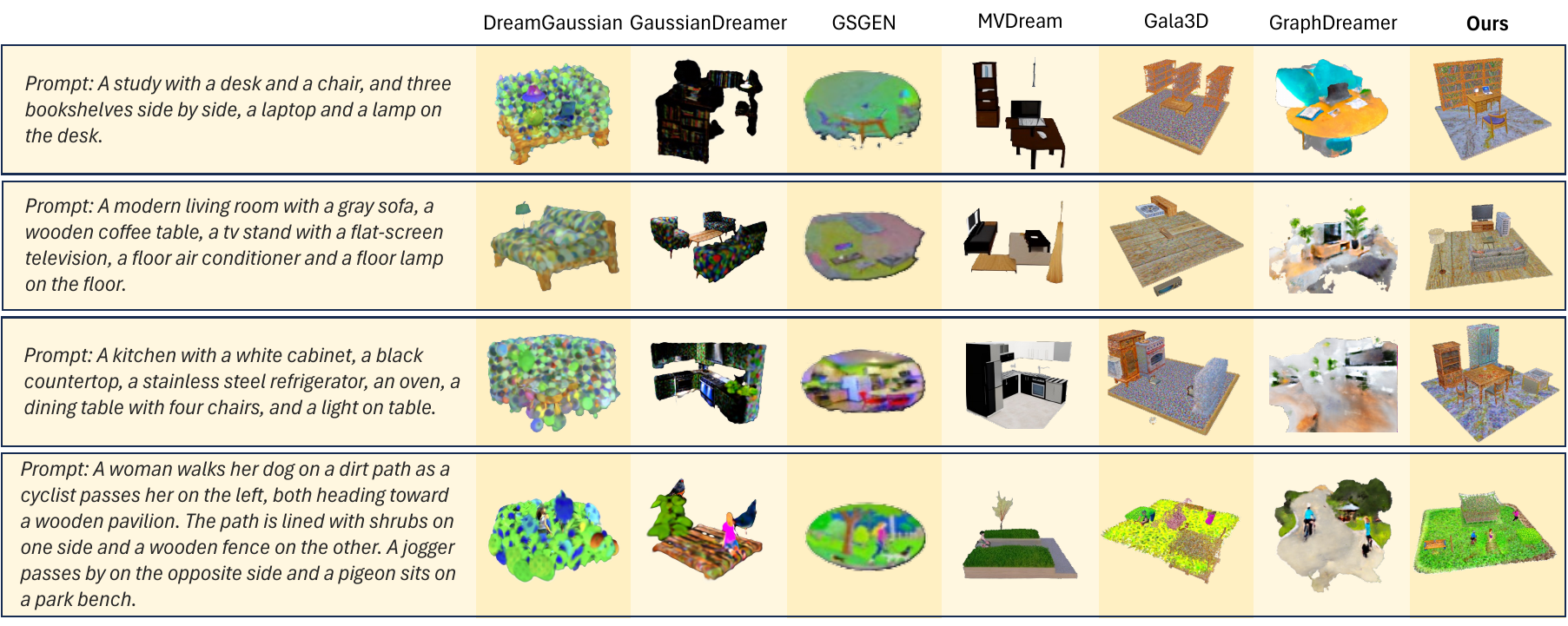}
   \caption{{Qualitative Reconstruction Results.} Compared to other methods, our approach produces high-quality reconstructions.}
   \label{fig:Qualitative}
 \end{figure*}

\begin{figure}[ht]
\centering
\includegraphics[width=0.95\linewidth]{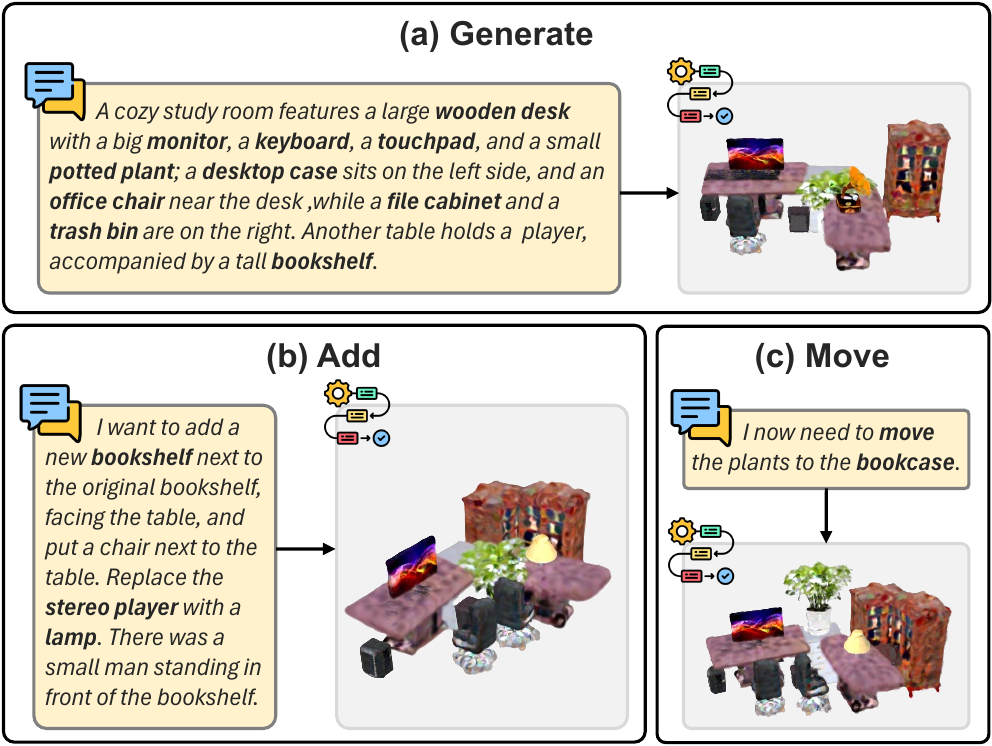}
\caption{{Scene Editing.} Our method can add, remove, or move objects based on the causal relationship between their placement.}
\label{fig:Editing}
\end{figure}

\subsection{Layout-guided Representation}
\label{3dgs_sds}
Based on causal reasoning for scene optimization and PID controller for object state adjustments, we obtain a structured layout, where each object is assigned a position based on its inferred dependencies. The layout provides center coordinates, and object sizes.

\paragraph{Object Representation}
Each object is optimized independently using MVDream \cite{mvdream} or Zero123 \cite{zero123} with Score Distillation Sampling (SDS):
\begin{equation}
    \nabla_{G_{\text{obj}}} L = \mathbb{E}_{\epsilon, \eta} \left[ \lambda_{\text{obj}} (\epsilon_{\phi} (I_{\text{obj}}; t, \beta, \eta) - \epsilon) \frac{\partial I}{\partial G_{\text{obj}}} \right],
\end{equation}
where $ G_{\text{obj}} $ represents object parameters, $ I_{\text{obj}} $ is the rendered object image, and $ t, \beta, \eta $ correspond to time step, camera parameters, and noise conditioning, respectively.

\paragraph{Scene Representation}
Simply placing objects directly in the scene makes it difficult to maintain overall coherence. To address this, we optimize the entire scene using Stable Diffusion \cite{rombach2022high} with SDS, ensuring consistency in object interactions. The full-scene optimization follows as:
\begin{equation}
    \nabla_{G_{\text{scene}}} L = \mathbb{E}_{\epsilon, p} \left[ \lambda_{\text{scene}} (\epsilon_{\phi} (I_{F_\text{scene}}; t, \beta, p) - \epsilon) \frac{\partial I}{\partial G_{\text{scene}}} \right],
\end{equation}
where $p$ represents the description of the entire scene, $ G_{\text{scene}} $ represents the global scene parameters, $ I_{F_\text{scene}} $ is the rendered full-scene image, and $ F_\text{scene} $ encodes the structured layout with 3D Gaussian properties and spatial parameters for all objects:
\begin{equation}
    F_\text{scene} = \{ \mathcal{G}_i, x_i, y_i, z_i, s_i\}_{i=1}^{N},
\end{equation}
where $ \mathcal{G}_i $ represents the 3D Gaussian attributes of object $ i $, while $ (x_i, y_i, z_i) $ denote the center coordinates, and $ s_i $ define the object scale. This joint optimization ensures that objects are not only individually refined but also integrated into a spatially coherent and semantically meaningful scene.

\section{Experiments}

\begin{table*}[t]
\centering

\scalebox{0.8}{\parbox{1.15\textwidth}{ 
\renewcommand{\arraystretch}{1} 
\begin{tabular}{@{} l c | c c c | c c c c c c @{}} 
\hline
\hline
\textbf{Methods} & \textbf{Represent.} & \textbf{ViT-B/32} & \textbf{ViT-L/14} & \textbf{ViT-bigL/14} & \textbf{Gemini} & \textbf{GPT-4o} & \textbf{Claude} & \textbf{Qwen} & \textbf{GLM} \\
\hline
MVDream~\cite{mvdream} & Nerf &24.30 & 18.34 & 18.16 & 1.7 & 4.1 & 2.7 & 1.6 & 3.6 \\
GraphDreamer~\cite{graphdreamer} & Nerf & 21.30 & 19.96 & 20.57 & 1.5 & 2.8 & 3.0 & 6.6 & 2.6 \\
\cmidrule(lr){1-10}
DreamGaussian~\cite{dreamgs} & 3DGS & 15.78 & 10.22 & 10.41 & 0.6 & 1.2 & 1.8 & 3.0 & 0.3 \\
GaussianDreamer~\cite{gsdreamer} & 3DGS & 20.57 & 18.04 & 18.72 & 0.8 & 1.8 & 1.7 & 5.2 & 1.0 \\
GSGen~\cite{gsgen} & 3DGS & 17.32 & 12.11 & 14.22 & 1.0 & 2.5 & 3.2 & 2.6 & 1.0 \\
GALA3D~\cite{gala3d} & 3DGS & 22.29 & 16.68 & 17.60 & 1.6 & 3.5 & 2.6 & 4.3 & 0.6 \\
\textbf{Ours} & \textbf{3DGS} & \textbf{25.90} & \textbf{20.86} & \textbf{21.31} & \textbf{2.8} & \textbf{5.0} & \textbf{3.3} & \textbf{6.9} & \textbf{4.2} \\
\hline
\hline
\end{tabular}
\caption{{Comparison with additional metrics.} CLIP~\cite{clip} \& MLLMs~\cite{gemini,GPT4o,qwen,glm,Claude}.}
\label{tab:clip_mllm_extended}
}
}
\end{table*}

\begin{table*}[t]

\centering
\scalebox{0.85}{\parbox{1\textwidth}{
\renewcommand{\arraystretch}{1} 
\begin{tabular}{lcc|cccccc}
\hline
\hline
\textbf{Methods} & \textbf{LLM} & \textbf{MLLM} &  \textbf{ViT-B/32} & \textbf{ViT-L/14} & \textbf{ViT-bigL/14} & \textbf{Gemini} & \textbf{GPT-4o} & \textbf{Claude} \\
\hline
w/o Causal Reasoning & - & - & 25.55 & 22.27 & 23.81 & 1.78 & 4.06 & 3.19  \\
DeepSeek(distilled) & R1-8b & LLava-34b & 27.48 & 23.05 & 24.88 & 2.39 & 4.83 & 3.75  \\
DeepSeek(distilled) & R1-14b & LLava-34b & \textbf{28.17} & 23.46 & 25.80 & 2.44 & 5.13 & 3.64\\
DeepSeek & R1 & 4o & 28.05 & 23.81 & 25.95 & \textbf{2.69} & 5.47 & \textbf{3.89}   \\
GPT & 4o & 4o & 27.81 & \textbf{24.63} & \textbf{26.95} & 2.42 & \textbf{5.64} & 3.64  \\
\hline
\hline
\end{tabular}
\caption{{Ablation Studies on Knowledge Distillation.} Our experiments systematically evaluate the impact of knowledge distillation, while probing how original non-distilled models shape final performance outcomes.}
\label{tab:abl_causal}
}

}

\end{table*}

 \begin{figure*}[t]
   \centering
   \includegraphics[width=0.9\linewidth]{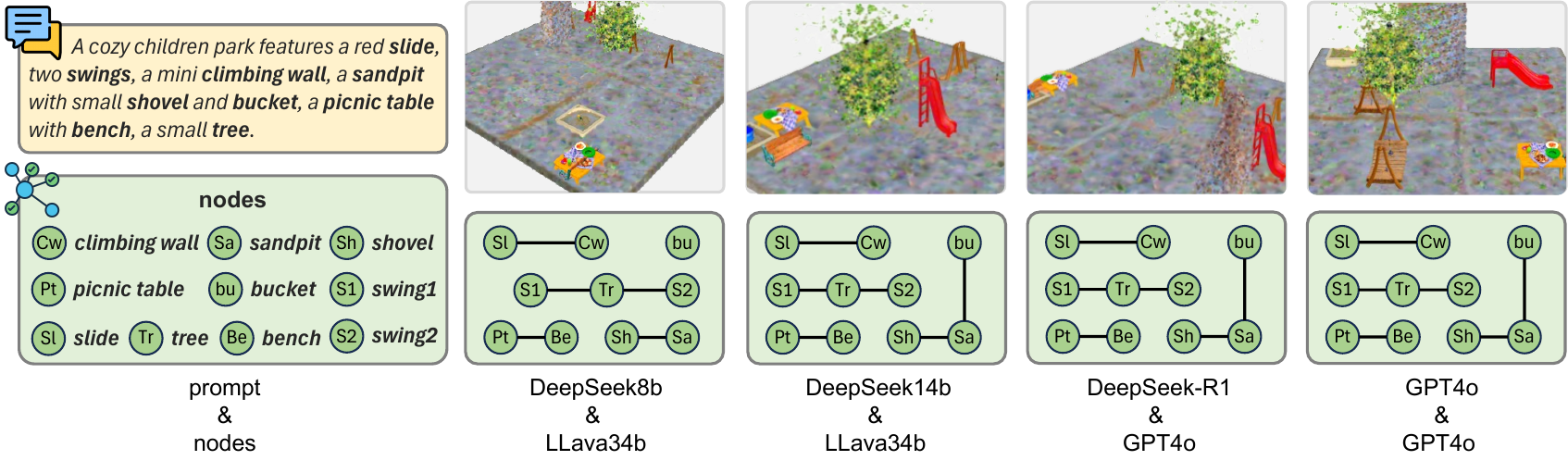}
   \caption{{Ablation Study of causal reasoning and adaptability.} The results show that causal reasoning enhances scene coherence, and experiments with distilled models demonstrate the robustness of our method.}
   \label{fig:causal}
 \end{figure*}

\begin{figure}[t]
   \centering
   \includegraphics[width=0.9\linewidth]{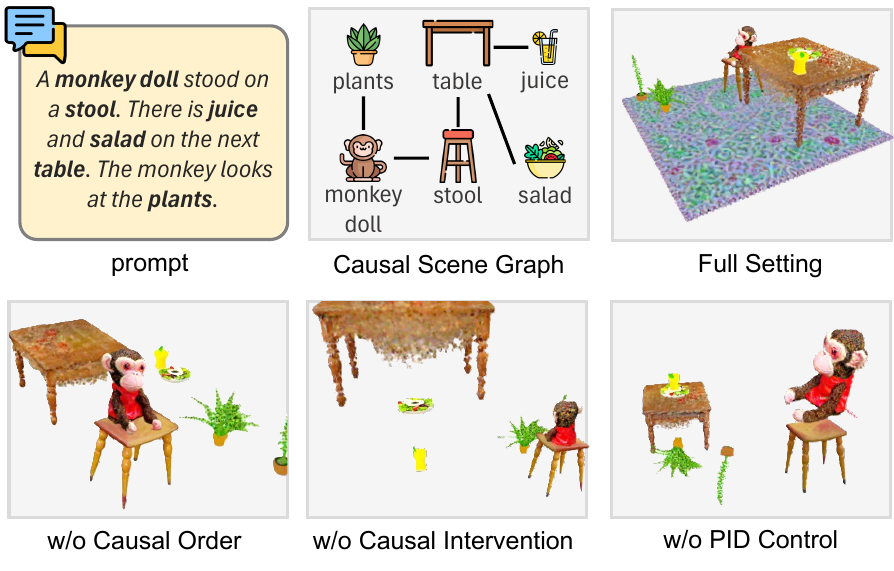}
   \caption{{Visual results of key Components.} The experiments validate the necessity of each component in our framework, highlighting their critical roles in ensuring coherent spatial relationships and physical plausibility}
   \label{fig:ablations}
 \end{figure}

\paragraph{Experimental Setup}
Our approach is implemented in PyTorch \cite{Pytorch}. We employ GPT-4o \cite{GPT4o} to generate the initial scene graph. To optimize the causal scene graph, we integrate DeepSeek \cite{guo2025deepseek} and GPT-4o, where DeepSeek facilitates strong chain-of-thought reasoning to refine causal relationships, while GPT-4o incorporates multimodal analysis to maintain consistency between textual descriptions and the visual layout. During causal graph optimization, we employ Point-E \cite{pointe} to render edge images and generate the spatial layout. During the object scale and position adjustment stage, we set $ k_p = 1, k_i = 0.00001, k_d = 5, \Delta = 0.02 $, and $ \gamma = 500$ for scale control, while for position control, we set $ k_p = 1, k_i = 0.00001, k_d = 5, \Delta = 0.4, $ and $ \gamma = 800 $. During Gaussian optimization and generation, we employ MVDream or Zero123 to refine individual objects and Stable Diffusion to optimize the overall scene, ensuring both object-level quality and scene-level coherence. 
All experiments were performed on an NVIDIA A100 GPU with 80GB memory.

\paragraph{Evaluation Metrics}
We evaluate our method using CLIP \cite{clip} Score and MLLMs \cite{gemini,GPT4o,qwen,Claude,glm} Score, comparing them quantitatively with baseline models. 
CLIP Score computes similarity by comparing the visual and textual embeddings extracted from the same CLIP model.
Additionally, we incorporate MLLMs Score, to further assess scene-object alignment and semantic coherence in the generated 3D representations.

\subsection{Comparison of Methods}
\paragraph{Quantitative Comparison}
We report quantitative results in Table~\ref{tab:clip_mllm_extended}. We assess our approach on the Text-to-3D task by comparing it with mainstream methods, including DreamGaussian \cite{dreamgs}, GaussianDreamer \cite{graphdreamer}, MVDream \cite{mvdream}, GSGen \cite{gsgen}, GALA3D \cite{gala3d}, and GraphDreamer \cite{graphdreamer}. 
To ensure a fair evaluation, we adopt CLIP Score, following prior works, to measure the alignment between generated images and their corresponding textual descriptions.
Given the inherent randomness in MLLMs, we utilize multiple MLLMs to assess the semantic consistency between generated scenes and input descriptions from different perspectives. 
Notably, higher CLIP or MLLM scores indicate better performance. By aggregating evaluations from CLIP and various MLLMs, our approach achieves the highest performance, demonstrating superior scene-object alignment and semantic coherence.

\paragraph{Qualitative Comparison}

We report qualitative comparisons on text-to-3D generation in Fig.~\ref{fig:Qualitative}. Compared to existing methods, our approach demonstrates superior spatial consistency and causal alignment. While prior methods primarily focus on single-object generation or data-driven scene synthesis, they often struggle with incorrect object relationships and spatial inconsistencies. In contrast, our method leverages causal order and intervention optimization to refine object interactions, ensuring that generated scenes adhere to real-world semantics and physical constraints. Additionally, our PID-based optimization maintains proportional accuracy, while diffusion-guided 3DGS refinement enhances overall rendering quality.

\subsection{Scene Editing}

As illustrated in Fig.~\ref{fig:Editing}, scene editing in our approach facilitates flexible and controllable modifications via text descriptions. 
LLMs translate user descriptions into layout transformations, such as adding, removing, or repositioning objects. The Layout-Guided Representation is subsequently optimized within the edited regions, maintaining stability in the unchanged areas.
Notably, our editing process accounts for the causal relationships that govern typical object placement, ensuring that modifications to position and scale align with real world spatial logic. This approach supports spatial adjustments, object interactions, and style modifications, offering a seamless and intuitive 3D scene editing experience grounded in causal reasoning.

In addition to generating 3D scenes from text descriptions, our method also supports text-image-based generation. 
Each node in the scene can receive image inputs, which guide the output of the node, enabling image-to-3D conversion. 
This integration enables text and image modes to work synergistically, where text provides global semantic constraints while images inject local geometric priors, thereby enhancing both the realism and physical plausibility of the synthesized scenes.

\subsection{Ablation Study}
\paragraph{Adaptability and Robustness}
Our framework incorporates localized, distilled LLMs and MLLMs, facilitating lightweight deployment while maintaining high performance across varied computational environments. As shown in Fig.~\ref{fig:causal} and Table~\ref{tab:abl_causal}, we assess multiple model configurations, demonstrating that our causal graph-based framework effectively captures object relationships and spatial dependencies. Leveraging a diverse set of models, our approach ensures robustness and stability in scene generation under varying computational constraints. Moreover, our framework efficiently adapts to varying input complexities, ensuring consistent spatial reasoning across diverse scenarios.

\paragraph{Causal Reasoning} 
To evaluate the impact of causal order and causal intervention on scene generation, we conduct ablation studies comparing scene graphs constructed from standard LLM parsing with those refined using causal reasoning. As shown in Fig.~\ref{fig:causal}, Fig.~\ref{fig:ablations} and Table~\ref{tab:abl_causal}, removing the causal order mechanism leads to missing essential object relationships, resulting in incomplete or incorrect connections.  Without Causal intervention to validate edges, the scene graph retains erroneous relationships, causing misaligned objects, floating placements, and unrealistic spatial arrangements. Integrating causal ordering and causal intervention validation enhances spatial consistency, object interactions, and physical plausibility.
These results highlight the need for causal reasoning in ensuring well-structured 3D representations.

\paragraph{PID Controller}

The PID controller regulates node attributes (object scales and positions) to maintain spatial consistency and enhance structural precision. As shown in Fig.~\ref{fig:ablations}, proportional optimization without PID control relies on fixed value updates, which are highly sensitive to erroneous LLM scores. These errors often propagate as perturbations in the system, causing fluctuations, directional misalignment, or even inverted object orientations—ultimately destabilizing scene layouts. In contrast, PID-based refinement addresses this limitation through a dynamic error-correction mechanism: the proportional term responds to immediate discrepancies, the integral term compensates for accumulated historical errors, and the derivative term anticipates abrupt changes. This multi-component control strategy effectively dampens noise from LLM evaluations, enabling smoother convergence. Through multi-term error compensation, PID dynamically regulates node attributes to generate physically coherent scenes in the case of abnormal LLM output.

\section{Conclusion}
In this paper, we proposed CausalStruct, a causal-driven framework for 3D scene generation, integrating causal reasoning,  PID control, and Diffusion refinement. By leveraging LLMs and MLLMs, our method constructs a causal scene graph, ensuring that object relationships align with real-world semantics and physical constraints. 
Moreover, our approach adapts to varying scene complexities, ensuring stable optimization across different generation tasks. 
Through PID control, we maintain proportional accuracy and spatial consistency, while 3DGS with SDS optimization enhances object fidelity and rendering quality. Experimental results show that CausalStruct improves scene composition, object interactions, and multi-view consistency, generating structured and semantically coherent 3D scenes. Our work demonstrates the potential of causal reasoning and PID control in 3D generation.

\bibliography{aaai2026}

\clearpage

\makeatletter
\@ifundefined{isChecklistMainFile}{
  \newif\ifreproStandalone
  \reproStandalonetrue
}{
  \newif\ifreproStandalone
  \reproStandalonefalse
}
\makeatother

\ifreproStandalone
\documentclass[letterpaper]{article}
\usepackage[submission]{aaai2026}
\setlength{\pdfpagewidth}{8.5in}
\setlength{\pdfpageheight}{11in}
\usepackage{times}
\usepackage{helvet}
\usepackage{courier}
\usepackage{xcolor}
\usepackage[hyphens]{url}  
\usepackage{graphicx} 
\urlstyle{rm} 
\def\UrlFont{\rm}  
\usepackage{natbib}  
\usepackage{caption} 
\usepackage{amsfonts} 
\usepackage{amsmath}
\usepackage{amssymb}
\usepackage{booktabs} 
\usepackage{graphicx} 
\frenchspacing  

\usepackage{algorithm}
\usepackage{algorithmic}

%
\usepackage{newfloat}
\usepackage{listings}
\usepackage{booktabs}    
\usepackage{array}        
\usepackage{graphicx}     
\usepackage{siunitx}     
\usepackage{caption}      

\begin{document}
\fi
\setlength{\leftmargini}{20pt}
\makeatletter\def\@listi{\leftmargin\leftmargini \topsep .5em \parsep .5em \itemsep .5em}
\def\@listii{\leftmargin\leftmarginii \labelwidth\leftmarginii \advance\labelwidth-\labelsep \topsep .4em \parsep .4em \itemsep .4em}
\def\@listiii{\leftmargin\leftmarginiii \labelwidth\leftmarginiii \advance\labelwidth-\labelsep \topsep .4em \parsep .4em \itemsep .4em}\makeatother

\setcounter{secnumdepth}{0}
\renewcommand\thesubsection{\arabic{subsection}}
\renewcommand\labelenumi{\thesubsection.\arabic{enumi}}

\newcounter{checksubsection}
\newcounter{checkitem}[checksubsection]

\newcommand{\checksubsection}[1]{%
  \refstepcounter{checksubsection}%
  \paragraph{\arabic{checksubsection}. #1}%
  \setcounter{checkitem}{0}%
}

\newcommand{\checkitem}{%
  \refstepcounter{checkitem}%
  \item[\arabic{checksubsection}.\arabic{checkitem}.]%
}
\newcommand{\question}[2]{\normalcolor\checkitem #1 #2 \color{blue}}
\newcommand{\ifyespoints}[1]{\makebox[0pt][l]{\hspace{-15pt}\normalcolor #1}}

\section{Clarify of Causal Reasoning}
In the introduction and related works, we define "Causal Reasoning (CR)" as explicitly modeling relationships between objects (e.g., spatial or functional relationships) while excluding irrelevant variables to address ambiguity in object placement during scene generation. This aligns with the definition of CR in LLMs \cite{xiong2024improving}.  
In image generation, consistent with Image Content Generation with Causal Reasoning \cite{li2024image}, CR in generative tasks aims to structure latent relationships between objects, compensating for missing details of objects during the generation process.

\section{Details of Bayesian Edge Estimation}
In our Bayesian Edge Estimation framework, we assume that each edge in the causal order edge set \( \mathcal{E}_{\text{order}} \) is independent given \( e_{i,j} \). 
However, edges may remain not independent in scenarios. The PID controller effectively addresses these dependencies by dynamically adjusting its parameters, ensuring comprehensive spatial coordination and precision across the framework.
This assumption allows us to apply the product rule for independent events in probability theory.
we compute the probability of the entire edge set occurring under the condition \( e_{i,j} \), referring to formula 5 in the main text. Under the condition that \(e_{i,j} \) is valid, the probability of each side can be calculated separately, without being affected by the presence of other sides in the set. This simplifies the calculation and allows us to decompose the joint probability into the product of the individual probabilities:
\begin{equation}
    p(\mathcal{E}_{\text{order}} \mid e_{i,j}) = p(e_{a,b} \mid e_{i,j}) p(e_{c,d} \mid e_{i,j}) \dots
\end{equation}
This calculation mode is particularly advantageous in LLM-based probability estimation, as it allows each edge to be evaluated independently without the need to explicitly model dependencies between them.

\section{Details of PID controller}

In our framework, causal reasoning and PID control operate as two distinct stages to ensure accurate scene reconstruction. Causal reasoning is first used to generate an initial object placement by inferring spatial dependencies from textual descriptions. However, these placements lack precise spatial alignment, leading to minor positional errors, intersections, or unrealistic gaps between objects.

To refine these placements, we employ a PID controller, which fine-tunes object positions and scales to ensure that the spatial configuration remains physically plausible. The PID controller takes the initial positions inferred from causal reasoning and iteratively adjusts them to minimize spatial discrepancies.

\begin{figure}[t]
   \centering
   \includegraphics[width=\linewidth]{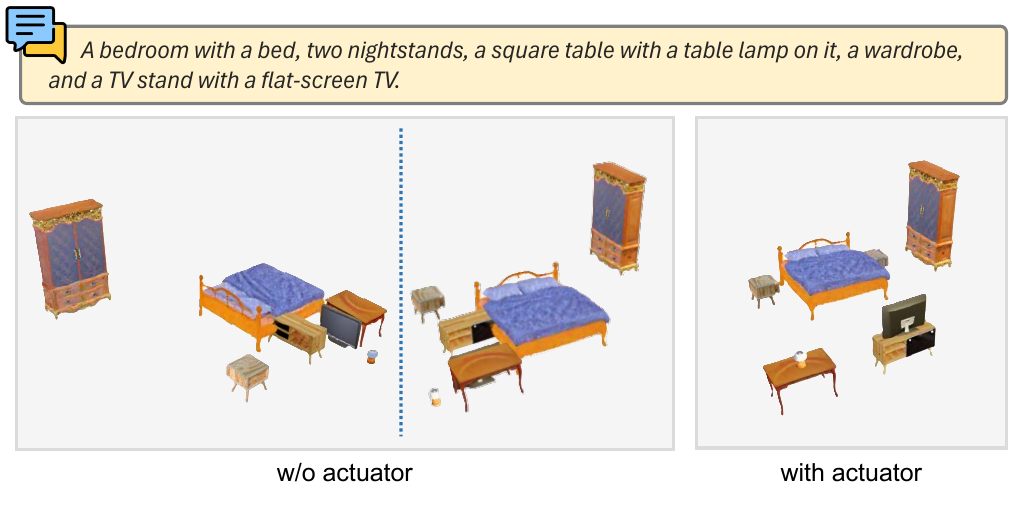}
   \caption{{Visual results of actuator.} }
   \label{fig:actuator01}
 \end{figure}

\begin{algorithm}[t]
\caption{PID Optimization with Iterative Loop.}
    \textbf{Input} p: Text prompt \\
    \textbf{Input} $\alpha$: Target score \\
    \textbf{Input} $\epsilon$: Error tolerance  \\
    \textbf{Input} $N$: Max iterations

\begin{algorithmic}[1]

    \STATE Initialize $\wp_{\text{current}}$, $E \gets 0$, $e_{\text{prev}} \gets 0$, $i \gets 0$
    \REPEAT
        \STATE $I_{edge} \gets render(\wp_{\text{current}})$ \COMMENT{Generate image}
        \STATE $s \gets \text{MLLM}\left(I_{edge}, \text{p}\right)$  \COMMENT{Generate score}
        \STATE $e \gets \alpha - s$ \COMMENT{Compute error}
        \STATE $E \gets E + e$ \COMMENT{Update integral}
        \STATE $d \gets e - e_{\text{prev}}$ \COMMENT{Compute derivative}
        \STATE $u \gets K_p \cdot e + K_i \cdot E + K_d \cdot d$ \COMMENT{PID control signal}
        \STATE $\Gamma \gets \Delta \cdot \tanh\left( u / \gamma \right)$ \COMMENT{Actuator output}
        \STATE $\wp_{\text{current}} \gets \wp_{\text{current}} + \Gamma$ \COMMENT{Update attribute}
        \STATE $e_{\text{prev}} \gets e$
        \STATE $i \gets i + 1$
    \UNTIL{$|e| \leq \epsilon$ \textbf{or} $i \geq N$}
 \RETURN $\wp_{\text{current}}$
\end{algorithmic}
\end{algorithm}

\paragraph{Why Use an Actuator?}
A key challenge in applying PID adjustments is preventing instability and overcorrection. 
As illustrated in Fig.\ref{fig:actuator01}, directly applying the PID output can lead to abrupt jumps, oscillations, or physically unrealistic movements, especially when the error signal fluctuates. To mitigate this, we introduce an actuator that smooths the adjustment process, ensuring controlled and realistic modifications to object placement.

Actuators are commonly used in control systems to translate control signals into gradual, physically constrained movements \cite{crowe2005pid}. In our case, the actuator controls position and scale adjustments, mitigating abrupt, excessive corrections that may introduce additional inconsistencies.

\paragraph{Why Use tanh as the Actuator?}
To implement the actuator, we use a hyperbolic tangent (tanh) function to smoothly transform the control signal into practical spatial adjustments. The choice of $ \tanh $ offers several advantages. 
\begin{itemize}
\item     \textbf{Saturation Effect:} The output of $ \tanh(x) $ is bounded between $-1$ and $1$, ensuring that extreme PID outputs do not result in excessively large movements, which could cause objects to shift too abruptly. 
\item     \textbf{Smooth Transitions:} Unlike a linear function, $ \tanh $ produces gradual transitions, which is essential for fine-tuning positions without introducing jerky or unnatural motion. 
\item    \textbf{Damping Small Adjustments:} For small error values, $ \tanh $ behaves approximately linearly, allowing precise micro-adjustments, while for large errors, it naturally limits the adjustment size, preventing excessive corrections.
    
\end{itemize}

\section{Compared with scene generation methods}
We further evaluate our approach against recent methods in compositional scene generation, including GraphDreamer\cite{graphdreamer}, GALA3D\cite{gala3d}, LI3D\cite{li3d}, and CompoNeRF\cite{componerf}. Detailed comparisons of GraphDreamer and GALA3D are presented in the main text. GALA3D utilizes a LLM to generate layouts. However, its direct layout inference lacks reasoning about object placement logic, often leading to physical constraint violations, such as objects floating in mid-air. GraphDreamer employs a graph-based structure to model inter-object relationships. 
\begin{figure}[t]
   \centering
   \includegraphics[width=0.9\linewidth]{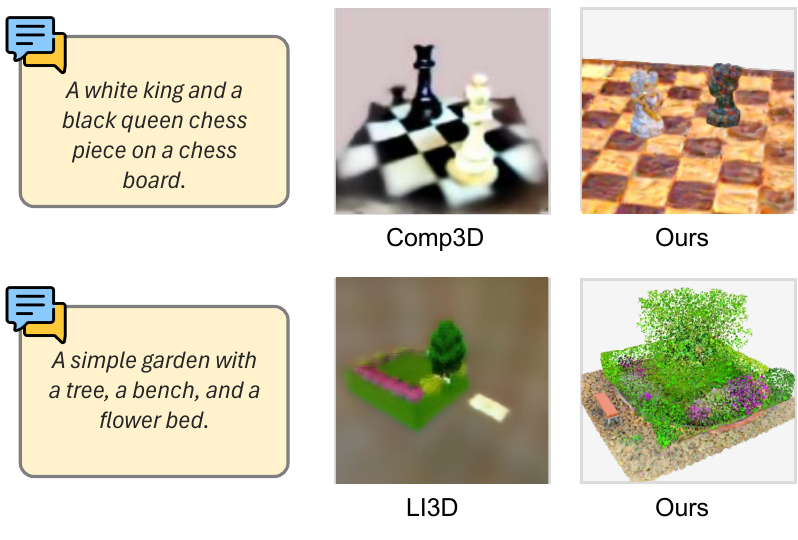}
   \caption{{Visual results of compositional scene generation methods.} }
   \label{fig:appendix01}
 \end{figure}
Despite this, experimental results demonstrate that its performance degrades significantly when generating complex scenes or environments with a large number of objects.
Since LI3D and CompoNeRF are not open-sourced, our comparison relies on the examples provided in their respective papers. As illustrated in Fig.\ref{fig:appendix01} , our method outperforms these approaches in terms of visual clarity and offers more precise scene control, enabling intuitive and interactive editing tailored to user specifications.

\section{Failure case}
While our LLM-based spatial evaluation is effective, distilled models with smaller architectures often struggle to establish certain relational edges and accurately adjust object positions due to reduced precision. This results in increased randomness in spatial adjustments and reduced accuracy in scene refinements. To mitigate this, we propose fine-tuning on scene layout datasets using contrastive learning and multi-view consistency constraints to enhance relational reasoning and positional accuracy.

\begin{table*}[ht]
\centering

\label{tab:causal_prompt}
\begin{tabular}{|p{0.95\linewidth}|}
\hline
\textbf{Causal Order Prompt} \\ 
\hline
\textbf{You are an expert in computer graphics, computer vision, causal analysis, and scene design.} \\
You will be provided with a scene layout graph containing objects (nodes) and their spatial relationships (edges). Your task is to analyze and refine this graph using physical constraints and causal reasoning. Follow these guidelines precisely: \\

\textbf{1. Allowed Spatial Relations:} \\
- All nodes need to have an edge. \\
- Use only the following words to describe connections: \{above, under, in, on, front, left, right, corner, behind, left\_front, right\_front, left\_back, right\_back, left\_on, right\_on\}. \\
- Only one word must be selected per edge. \\

\textbf{2. Causal Reasoning \& Edge Completion:} \\
- If two objects are closely related in real-world use but are not connected in the input graph, infer the missing edge and add it (e.g., add [``lamp", ``on", ``table"] if missing, but do not add ``floor"). \\
- Ensure causal flow integrity: All edges must form a directed acyclic graph following causal order. \\

\textbf{3. Causal Order Principles:} \\
- Objects follow a causal flow: \( obj_2 \) comes \textbf{after} \( obj_1 \) if \( obj_1 \)'s placement depends on \( obj_2 \) (e.g., [``cup", ``on", ``table"]). \\
- \textbf{Causal Rule:} If \( obj_1 \) depends on \( obj_2 \), reverse edge and adjust relation. Example: [``table", ``under", ``cup"] → [``cup", ``on", ``table"]. \\
- \textbf{Size Rule:} Larger objects should be \( obj_2 \). Transform [``laptop", ``left\_on", ``mouse"] → [``mouse", ``right\_on", ``laptop"]. \\
- Causal order takes priority over size when they conflict (e.g., [``TV", ``on", ``stand"], even if TV is larger). \\

\textbf{Output Format:} The output must contain the following content: \\
\texttt{<Answer>edges = [[obj\_1, word\_1, obj\_2], [obj\_2, word\_4, obj\_3], ...]</Answer>}. \\

\textbf{Example Corrections:} \\
\textbf{Input:} [['laptop','left','mouse'], ['cup','under','table']] \\
\textbf{Output:} edges = [['mouse','right','laptop'], ['cup','on','table']] \\
\hline
\end{tabular}
\caption{Prompt used for causal order inference in our framework. The LLM is guided to ensure spatial constraints, causal consistency, and object dependencies.}
\end{table*}

\begin{table*}[t]
\centering
\label{tab:causal_intervention_prompt}
\begin{tabular}{|p{0.95\linewidth}|}
\hline
\textbf{Causal Intervention Prompt} \\ 
\hline
\textbf{I will provide an image of a scene.} \\

The image depicts \texttt{\{candidate\_edge.create\_prompt()\}} as part of the scene described as: \texttt{\{prompt\}}. \\
In this scene, object '\texttt{\{obj\_names[0]\}}' is currently labeled as '\texttt{\{candidate\_edge.edge\_name\}}' relative to '\texttt{\{obj\_names[1]\}}'. \\

\textbf{Task:} \\
- Assess whether the given spatial relationship complies with physical laws and real-world scene consistency. \\
- Based on the provided image and object interactions, determine the validity of the relation using the following criteria: \\
  - Gravity \& Support: Objects must adhere to realistic physical constraints (e.g., smaller objects should rest on larger ones, and unsupported objects should not float). \\
  - Spatial Positioning: The labeled relationship should match common spatial arrangements (e.g., a chair should be under a table, not above it). \\
  - Functional Affordance: Objects should maintain plausible real-world functionality (e.g., a monitor should be on a desk, not inside it). \\

\textbf{Decision Guidelines:} \\
- If the relationship is valid, return 'keep'. \\
- If the relationship is incorrect but fixable, return 'modify' and suggest a new relation from the predefined set: \texttt{\{candidate\_relations\}}. \\

\textbf{Output Format:} \\
Provide the response strictly in JSON format as follows: \\
\texttt{\{
    "action": "keep" | "modify",
    "updated\_relation": "new\_relation" 
\}} \\

\hline
\end{tabular}
\caption{Prompt used for causal intervention in our framework. The LLM evaluates scene images across multiple perspectives to assess the correctness of object relationships, ensuring alignment with real-world physics, and spatial reasoning.}
\end{table*}

\begin{table*}[t]
\centering

\label{tab:scale_evaluation_prompt}
\begin{tabular}{|p{0.95\linewidth}|}
\hline
\textbf{Scale Evaluation Prompt} \\ 
\hline
\textbf{I will provide an image of a scene.} \\

\textbf{Object Dimensions:} \\
- The dimensions (length, width, height) of \texttt{\{obj\_names[0]\}} in the real world: \\
  - Length: \texttt{\{length0\}} cm, Width: \texttt{\{width0\}} cm, Height: \texttt{\{height0\}} cm. \\
- The dimensions (length, width, height) of \texttt{\{obj\_names[1]\}} in the real world: \\
  - Length: \texttt{\{length1\}} cm, Width: \texttt{\{width1\}} cm, Height: \texttt{\{height1\}} cm. \\

\textbf{Task:} \\
- Evaluate the relative scale of the object \texttt{\{obj\_names[0]\}} compared to \texttt{\{obj\_names[1]\}}. \\
- The scale of \texttt{\{obj\_names[1]\}} is assumed to be correct, but \texttt{\{obj\_names[0]\}} may have scaling inconsistencies in the scene with \texttt{\{edge.create\_prompt()\}}. \\

\textbf{Evaluation Criteria:} \\
Scale Comparison: Does \texttt{\{obj\_names[0]\}} appear appropriately scaled relative to \texttt{\{obj\_names[1]\}}? \\
   - Consider the effect on scene composition, ensuring it is neither too large nor too small. \\

\textbf{Scoring System:} \\
- A score from -100 to 100 is assigned based on scale consistency: \\
  - Score close to 0: The scale of \texttt{\{obj\_names[0]\}} is appropriate. \\
  - Positive score: \texttt{\{obj\_names[0]\}} is too large compared to \texttt{\{obj\_names[1]\}}, disrupting scene balance. \\
  - Negative score: \texttt{\{obj\_names[0]\}} is too small, making it insignificant in the scene. \\

\textbf{Output Format:} \\
Provide only the result in the following format, with no additional text: \\
\texttt{<Answer>The score is: X</Answer>}, where X is the evaluated score. \\
For example, output: \texttt{<Answer>The score is: 25</Answer>}. \\

\hline
\end{tabular}
\caption{Prompt used for scale evaluation and PID-based optimization in our framework. The LLM assesses object-relative scaling.}
\end{table*}

\begin{table*}[t]
\centering
\label{tab:position_evaluation_prompt}
\begin{tabular}{|p{0.95\linewidth}|}
\hline
\textbf{Spatial Position Evaluation Prompt} \\ 
\hline
\textbf{I will send you a sentence and images of a scene.} \\

\textbf{Scene Description:} \\
The image shows a scene of \texttt{\{edge.create\_prompt()\}}. \\

\textbf{Evaluation Task:} \\
- The position of object \texttt{\{obj\_names[1]\}} is correct. \\
- Object \texttt{\{obj\_names[0]\}} may be misplaced in the scene. Evaluate its spatial deviation along three axes. \\

\textbf{Scoring Criteria:} \\
Assign a score from -100 to 100 along each axis: \\
1. Left-Right (X-Axis): \\
   - Positive score: Too close to \texttt{\{obj\_names[1]\}}. \\
   - Negative score: Too far from \texttt{\{obj\_names[1]\}}. \\
2. Forward-Backward (Y-Axis): \\
   - Positive score: Too close to \texttt{\{obj\_names[1]\}}. \\
   - Negative score: Too far from \texttt{\{obj\_names[1]\}}. \\
3. Up-Down (Z-Axis):  \\
   - Positive score: Too high above \texttt{\{obj\_names[1]\}}. \\
   - Negative score: Too low below \texttt{\{obj\_names[1]\}}. \\

\textbf{Output Format:} \\
\texttt{<Answer>The score-1 is: XX. The score-2 is: YY. The score-3 is: ZZ</Answer>} \\

\hline
\end{tabular}
\caption{Prompt used for spatial position evaluation. The LLM provides axis-aligned position corrections based on multi-view scene analysis.}
\end{table*}

\ifreproStandalone
\end{document}
\fi

\end{document}